\documentclass[longauth]{aa} % for the long lists of affiliations 
\usepackage{graphicx}
\usepackage{txfonts}
\begin{document}

\title{MAGIC observations of the giant radio galaxy M87 in a low-emission state between 2005 and 2007}


\author{
 J.~Aleksi\'c\inst{1} \and
 E.~A.~Alvarez\inst{2} \and
 L.~A.~Antonelli\inst{3} \and
 P.~Antoranz\inst{4} \and
 M.~Asensio\inst{2} \and
 M.~Backes\inst{5} \and
 J.~A.~Barrio\inst{2} \and
 D.~Bastieri\inst{6} \and
 J.~Becerra Gonz\'alez\inst{7,}\inst{8} \and
 W.~Bednarek\inst{9}
% ...
}
\institute { IFAE, Edifici Cn., Campus UAB, E-08193 Bellaterra, Spain
 \and Universidad Complutense, E-28040 Madrid, Spain
 \and INAF National Institute for Astrophysics, I-00136 Rome, Italy
 \and Universit\`a  di Siena, and INFN Pisa, I-53100 Siena, Italy
 \and Technische Universit\"at Dortmund, D-44221 Dortmund, Germany
 \and Universit\`a di Padova and INFN, I-35131 Padova, Italy
 \and Inst. de Astrof\'{\i}sica de Canarias, E-38200 La Laguna, Tenerife, Spain
 \and Depto. de Astrof\'{\i}sica, Universidad de La Laguna, E-38206 La Laguna, Spain
 \and University of \L\'od\'z, PL-90236 Lodz, Poland
% ...
}
\date{Compiled 04 August 2011}

\abstract
  {We present the results of a long M87 monitoring campaign in very high energy $\gamma$-rays with the MAGIC-I Cherenkov telescope.}
  {We aim to model the persistent non-thermal jet emission by monitoring and characterizing the very high energy $\gamma$-ray emission of M87 during a low state.}
  {A total of 150\,h of data were taken between 2005 and 2007 with the single MAGIC-I telescope, out of which 128.6\,h survived the data quality selection. We also collected data in the X-ray and \textit{Fermi}--LAT bands from the literature (partially contemporaneous). } 
  {No flaring activity was found during the campaign. The source was found to be in a persistent low-emission state, which was at a confidence level of $7\sigma$. We present the spectrum between 100\,GeV and 2\,TeV, which is consistent with a simple power law with a photon index $\Gamma=2.21\pm0.21$ and a flux normalization at 300\,GeV of $(7.7\pm1.3) \times 10^{-8}$ TeV$^{-1}$ s$^{-1}$ m$^{-2}$. 
%The spectrum matches well the lower-energy points published by \textit{Fermi}, extending its energy range by four orders of magnitude, without evidence for modification in the power law index. !! 
The extrapolation of the MAGIC spectrum into the GeV energy range matches the previously published \textit{Fermi}--LAT spectrum well, covering a combined energy range of four orders of magnitude with the same spectral index. We model the broad band energy spectrum with a spine layer model, which can satisfactorily describe our data.}
  {}

\keywords{gamma-rays: galaxies --- individual: M87}
                            
\titlerunning{M87}
                             
\authorrunning{Aleksi\'c et al.}
                            
\maketitle
%________________________________________________________________

\section{Introduction}

%AGN

Current astrophysical observations suggest that the centre of almost every galaxy harbours a super-massive black hole (SMBH) with a mass of the order of $10^6 - 10^{10} M_\odot$. Depending on the evolutionary stage and the type of the galaxy, the amount of interstellar matter in the vicinity of the central black hole differs significantly. In about 1\% of the galaxies the gravitational energy released by matter falling onto the central SMBH is sufficiently high to make the  nucleus more luminous than the entire galaxy, hence the name of active galactic nuclei (AGN, see Urry \& Padovani 1995; Urry 2003; Vellieux 2003). In a small group of AGN, which are called radio-loud, mechanisms (mediated by the magnetic field) acting close to the vicinity of the SMBH are able to energise and collimate the flow of matter in the form of relativistic jets, which extend to distances of kpc up to Mpc from the centre. 
The exact mechanisms of this process, as well as the source of energy that accelerates particles to relativistic velocities, are not yet entirely understood (see for instance the Blandford--Payne scenario Blandford \& Payne 1982, or the Blandford--Znajek process Blandford \& Znajek 1977). 

Once ejected through the jet, the particles interact with each other and the surrounding magnetic and radiation fields producing photons with energies up to TeV (Marashi et al.\ 1992; Bloom \& Marscher 1996). The collimation of the jet and the relativistic boosting of the emitted radiation imply that relativistic jets appear brightest when they are inclined by a small angle towards the observer, as is the case of the sources known as \textit{blazars}. Owing to their large distances and the very low inclination of their jets to the line of sight, it is difficult to study the structure of these sources. However, observation of some nearby radio-loud AGN that have a higher jet inclination angle (such as M87) give us the opportunity to study jets in greater detail, since the angular resolutions of radio to X-ray observatories is sufficient to resolve their structures (Wilson \& Yang 2002).

%M87

M87 is a giant elliptical radio galaxy at the centre of the Virgo cluster at a distance of 16.7 Mpc (Mei et al.\ 2007). It is harbouring a SMBH with a mass of $(6.4 \pm 0.5)\times 10^9 $M$_\odot$ (Gebhardt \& Thomas 2009). The prominent jet of M87, first discovered in the visual observation of Curtis (1918), is inclined by $10^\circ - 45^\circ$ from our line of sight (Biretta, Sparks \& Macchetto 1999; Ly, Walker \& Junor 2007). The vicinity of M87 makes it possible to study different parts of the jet in detail. 

The emission of the compact nucleus as seen in GHz frequencies is variable and has a dimension of several hundred Schwarzschild radii. Throughout the inner section of the jet several brighter spots, so-called ``knots", are visible, often characterized by superluminal motion with apparent speeds reaching $\sim$6\,c (Biretta, Zhou, \& Owen 1995; Biretta, Sparks \& Macchetto 1999). The one closest to the nucleus (60\,pc), known as HST--1, contains blobs with superluminal motion, as revealed by radio VLBI observations (Cheung et al.\ 2007), and shows high variability of its optical and X-ray emission (e.g.\ Perlman et al.\ 2003, Harris et al.\ 2009). A closer inspection also indicates the existence of a counter-jet (Biretta, Sparks \& Macchetto 1999). Recently, \cite{Hada} used the core-shift effect detected in Very Long Baseline Array (VLBA) multi-frequency data between 2 and 43\,GHz to estimate that the central engine of M87 lies within a distance of 14--23 Schwarzschild radii of the radio core at 43\,GHz.

The first hint of very high energy (VHE, E$>$100\,GeV) $\gamma$-ray emission reported by the HEGRA collaboration (Aharonian et al.\ 2003) triggered extensive observations by the next generation Imaging Atmospheric Cherenkov telescopes (IACTs): The H.E.S.S. collaboration firmly established M87 as an emitter above 730\,GeV (Aharonian et al.\ 2006) and revealed flux variability on time-scales of two days, suggesting that the emission region of the $\gamma$-rays is very compact with a dimension similar to the Schwarzschild radius of the central black hole. VERITAS also detected VHE $\gamma$-radiation from M87 in 2007 (Acciari et al.\ 2008) above 250\,GeV and subsequently monitored the source during the following years (see e.g. Acciari et al.\ 2010).  The first reported detection of $\gamma$-ray emission from M87 by the Major Atmospheric Gamma-Ray Imaging Cherenkov (MAGIC) collaboration used data taken in 2008 during a flaring state (Albert et al.\ 2008b). These data were collected during a monitoring programme together with H.E.S.S., VERITAS and VLBA (Acciari et al.\ 2009).

Using the Large Area Telescope (LAT) on-board the Fermi Gamma-ray Space Telescope ({\it Fermi}--GST) high-energy $\gamma$-radiation above 100\,MeV (up to 30\,GeV) from M87 was detected (Abdo et al.\ 2009) during the first ten months of the mission. The spectral index of 2.26$\pm$0.13 is consistent with the VHE measurements and no significant variability was found.            

The large-scale jet primarily emits from radio to X-rays via the synchrotron mechanism. The processes by which the high-energy radiation is produced, as well as the regions where these processes take place is more debated and has been extensively studied in the past years, based especially on radio, optical, X-ray, and $\gamma$-ray observations. The two most promising high-energy emission regions are the nucleus and the knot ``HST-1". At X-rays, \textit{Chandra} has the angular resolution to separate the two components, showing a quite complex behaviour (Harris et al.\ 2009). Unfortunately, the angular resolution of $\gamma$-ray observations does not allow us to resolve the jet structure, but the observed variability can be used to constrain the size of the emission region by requiring that the variability time-scale is longer than the light travel time {\it R/c} ({\it R} is the typical radius of the emission region). Even if this information alone is (strictly speaking) unable to reveal the location of the emission, it can be correlated with multiwavelength data in which the source is resolved. For the particular case of M87, studies by \cite{MAGIC_flare} and \cite{Science} have thus favoured the nucleus of M87 as the most likely origin for the VHE $\gamma$-radiation (see also Abramowski et al.\ 2012 for a recent summary).

The discovery of high-energy $\gamma$-ray emission from M87 (and more generally from radio-galaxies) stimulated an intense theoretical work aimed at clarifying the mechanisms that produce the observed radiation.
As already anticipated, the rapid ($\sim$day) variability points to a very compact emission region, ruling out early models that invoked emission from the large-scale ($\sim$kpc) jet (Stawarz et al.\ 2003). Two main theoretical scenarios have been discussed in the literature: in the first one the emission is believed to arise in the close vicinity of the SMBH through electromagnetic mechanisms acting in the magnetosphere of the black hole (e.g.\ Neronov \& Aharonian 2007; Rieger \& Aharonian 2008; Levinson \& Rieger 2011). In the other class of models, the emission is postulated to originate in the innermost region of the jet, as with blazars. The models differ on the geometry of the emission regions and on the emission mechanism (e.g. Georganopoulos et al.\ 2005; Lenain et al.\ 2008;  Tavecchio \& Ghisellini 2008; Giannios et al.\ 2010; Barkov et al.\ 2010). In the latter class of models, multiple regions are generally assumed to fully reproduce the observed spectral energy distribution (SED).
In particular in the structured jet model of Ghisellini et al. (2005) and Tavecchio \& Ghisellini 2008 one envisages that the jet has an inner, faster (bulk Lorentz factor $\Gamma=10-15$) ``spine" surrounded by a slower ($\Gamma=3-4$) sheath. The radiative interaction between the two regions leads to an effective enhancement of the inverse Compton emission and the possibility to decelerate the spine through the "Compton drag" effect (Begelman \& Sikora 1987). This structure may account for the emission properties of both blazars (observed at small angles, for which the emission is dominated by the spine) and radio-galaxies (larger viewing angles and substantial emission from the layer). 
Another possibility is the involvement of electromagnetic cascades from relativistic electrons that propagate along the jet (Sitarek \& Bednarek; and Roustazadeh \& B\"ottcher 2010). Finally, other models assume emission from the jet but identify the emission region in the HST-1 knot, invoking extreme re-collimation of the flow to reproduce the fast variability (e.g. Stawarz et al.\ 2006; Bromberg \& Levinson 2009).

Whereas in the cases mention above the models have been applied to interpret high or flaring states of M87, a more detailed look into the characteristics of the source's lower emission levels is still lacking. In this paper we will focus on the analysis of long-term monitoring observations of M87 with MAGIC during a low-emission state between 2005-2007.

\section{Observations and data analysis}

The data were collected with the stand-alone MAGIC-I 17\,m diameter imaging atmospheric Cherenkov telescope situated in the Canary Island of La Palma at the Roque de los Muchachos Observatory at an altitude of 2200\,m above the sea level. The energy threshold of the standard trigger configuration for observations at low zenith angles is about 60\,GeV. The angular resolution is $\sim$0.1$^\circ$ on an event-to-event basis and the energy resolution above 150\,GeV is $\sim$25\% (see Albert et al.\ 2008a for details).

MAGIC started to observe M87 regularly in 2005 and continued to monitor it during several observing cycles until 2012 (see e.g. the recent announcement of an enhanced emission state in February 2010: Mariotti et al.\ 2010 and Abramowski et al.\ 2012). This paper focuses on data taken between March 2005 and June 2007 with the stand-alone MAGIC-I telescope. The results of the 2008 observing campaign, when M87 entered an intense high state, have been previously published (see Albert et al.\ 2008b and Acciari et al.\ 2009 for details). 

The MAGIC-I telescope underwent several significant upgrades during the long observation period. These upgrades specifically include that optical point spread function (PSF) of the 17\,m mirror dish was improved due to a better understanding of the automatic mirror control (AMC) from about 20\,mm at the beginning of 2005 to 13\,mm at the end of April 2005. In the subsequent observing time the telescope underwent several distinct mirror-focusing adjustments that resulted in a varying optical PSF between 13\,mm and 15\,mm. While the varying PSF has no significant effect on our angular resolution, it does affect our collection area, especially near the threshold. These effects have been taken into account in the Monte Carlo simulation.

The readout chain has been significantly enhanced during the observing time. While the data recorded until April 2006 used a 300\,MSample/s flash analogue-to-digital converter (FADC) system, subsequently a new 2\,GSample/s Mux-FADC system was installed in February 2007 for the whole MAGIC-I camera. During the transition time between April 2006 until February 2007 a splitter system was installed (and continuously used until today), which allowed splitting the signal between the Mux-FADC readout and the trigger electronics. Accordingly, the data discussed in this paper include three different readout stages: {\it i)} 300\,MS/s readout without splitters, {\it ii)} 300\,MS/s readout including splitters, and {\it iii)} 2\,GS/s readout with splitters. Events recorded with the 300\,MS/s system were stretched up to 6\,ns full width half maximum (FWHM) (10\,ns for the low gain) to ensure proper sampling. With the introduction of the new 2\,GS/s readout system this stretching as well as the high and low gain differentiation became unnecessary. 

MAGIC observations of M87 were performed in two different modes: ON-OFF, and wobble. In the ON-OFF mode the telescope is pointed directly at the source during the ON runs. The background is estimated with dedicated OFF runs, which are recorded separately in a field of view of the sky where no $\gamma$-ray signal is expected but the night sky background (NSB) and zenith angles match the ON observations. It is imperative that the telescope conditions are the same for the ON and the OFF data to avoid additional systematic effects. Accordingly, only OFF data taken during the same observing period as the ON data were used in this analysis. Only the 2005 observations of M87 were recorded in ON-OFF mode (see also Table 1). Data taken in the later years were collected in the so-called {\it wobble} mode (Fomin et al.\ 1994) where the telescope is not pointed directly at the source position, but 0.4$^{\circ}$ away.  Accordingly, wobble data do not require dedicated OFF runs since the background can be estimated from points in the camera that are equivalent to the source position. Those positions are equidistant from the camera centre and have sufficient separation from the source location so that the expected $\gamma$-ray signal does not spill over into the OFF region. 

Throughout this paper we have applied the same timing-based analysis to the entire dataset, as described in \cite{Mrk421}. After applying the standard calibration and signal extraction algorithms (Albert et al.\ 2008c), we have performed the so-called "time image cleaning" procedure (Aliu et al.\ 2009), which uses the strictly correlated nature of the Cherenkov light pulses (both in time and in space) to distinguish from spurious (randomly distributed) signals from NSB and electronic noise. The first step requires at least six photoelectrons in the core pixels and three photoelectrons in the boundary pixels of the images (Fegan 1997). Only pixels with at least two neighbouring pixels, whose photons arrive within less than 1.75\,ns time difference, survive the second cleaning step. The third step repeats the cleaning of the second step, but requires only one adjacent pixel within the 1.75\,ns time window.

We used the recipe of \cite{Hillas} to calculate image parameters such as WIDTH (the root mean squared of the shower image amplitude along the minor axis), LENGTH (the root mean squared of the shower image amplitude along the major axis) and SIZE (corresponding to the total light content of the image). The strong domination of air showers induced by charged cosmic-rays over $\gamma$-ray induced showers requires a strong event selection. This so-called "$\gamma$--hadron separation" uses a SIZE-dependent parabolic cut in AREA $=$ WIDTH x LENGTH x $\pi$ (Riegel et al.\ 2005). The cut parameters were optimised using contemporaneous Crab Nebula data. The analysis energy threshold (defined as the peak of the Monte Carlo energy distribution) corresponds to $\sim$140\,GeV.

The energy of the primary $\gamma$-ray events was estimated using the Random Forest regression method (Albert et al.\ 2008d) trained on Monte Carlo simulated $\gamma$-ray events. We used the DISP method (Fomin et al.\ 1994, Lessard et al.\ 2001) to reconstruct the arrival direction of the events, which was extended to include the shower timing information as described in \cite{Mrk421}.

Each period with different telescope conditions was first analysed separately with matching OFF-data (when needed) and Monte Carlo simulations from the corresponding period and finally combined after the $\gamma$--hadron separation. A stacked analysis of long-term MAGIC-I observations was first performed in \cite{stacking}. An excellent agreement between the published Crab Nebula spectrum and the stacked analysis was found, confirming the stability of the analysis over long time-scales. 

The entire observing campaign consists of 154.1\,h out of which 128.6\,h survived the quality selection cuts. This selection ensures that only data with good weather conditions and without any technical problems are used in the analysis. Table \ref{tab:obs.periods} gives an overview of how these datasets were distributed over the entire campaign. M87 culminates at $16^\circ$ when observed from La Palma and most of the data were taken with low zenith angles (below $30^\circ$ with some exceptions up to $45^\circ$).

%\begin{center}
%\begin{tabular}{| c | c | c | c | c |}
%  \hline
%  Obs period & Obs mode & read-out system & PSF & Obs time \\
%  2005a & On-Off & 300\,MS/s & 20\,mm & 6.5\,h \\ \hline
%  2005b & On-Off & 300\,MS/s & 13\,mm & 1.5\,h \\ \hline
%  2006a & Wobble & 300\,MS/s & 15\,mm & 53.4\,h \\ \hline
%  2006b & Wobble & 300\,MS/s + Splitters & 14\,mm & 26.0\,h \\ \hline
%  2007 & Wobble & 2\,GS/s + Splitters & 14\,mm & 42.8\,h \\ \hline
% \hline
%\end{tabular}
%\end{center}

\begin{table*}
\caption{List of the effective observing times obtained with each hardware setup. See text for details.}
\label{tab:obs.periods}
\centering
\begin{tabular}{ c c c c c c }
\hline \hline
Obs. period    
& Obs. date  
& Obs. mode  
& read-out system
& PSF  
& Obs. time  \\
\hline  
2005a & 2005-03-15 - 2005-04-10 & On-Off & 300\,MS/s & 20\,mm & 6.5\,h \\ 
2005b & 2005-05-07 - 2005-05-08 & On-Off & 300\,MS/s & 13\,mm & 1.5\,h \\ \hline
2006a & 2005-12-30 - 2006-04-03 & Wobble & 300\,MS/s & 15\,mm & 53.4\,h \\ 
2006b & 2006-05-23 - 2006-12-31 & Wobble & 300\,MS/s + Splitters & 14\,mm & 26.0\,h \\ \hline
2007  & 2007-02-13 - 2007-05-19 & Wobble & 2\,GS/s + Splitters & 14\,mm & 42.8\,h \\
\hline
\end{tabular}
\end{table*}

\section{MAGIC results}

\begin{figure}
\includegraphics[angle=270,width=0.49\textwidth]{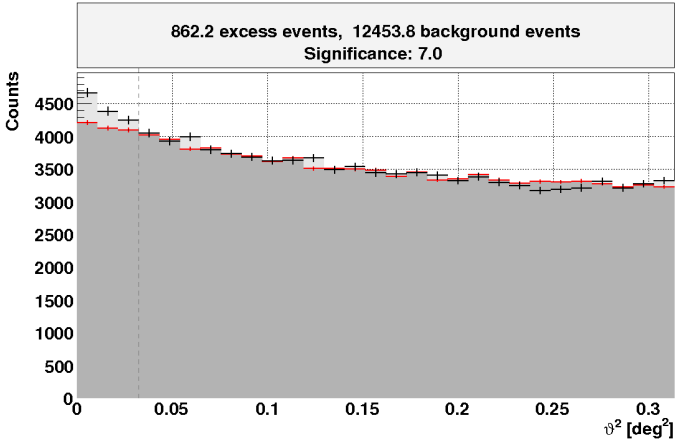}
%\caption{$\vartheta^2$ distribution of the combined 128.6\,h MAGIC observations between 2005 and 2007. $\vartheta^2$ is the squared angular distance between the source position and the reconstructed arrival direction of each event. The grey shaded area below the red crosses is the OFF data-sample, while the black crosses correspond to the ON data. The excess is point-like compared to the PSF with a significance of 7 standard deviations (calculated with formula 17 of \citep{Li \& Ma 1983}). }
%\caption{$\vartheta^2$ distribution of the combined 128.6\,h MAGIC observations between 2005 and 2007. $\vartheta^2$ is the squared angular distance between the source position and the reconstructed arrival direction of each event. The grey shaded area below the red crosses is the OFF data-sample, while the black crosses correspond to the ON data. The excess is point-like compared to the PSF with a significance of 7 standard deviations (calculated with formula 17 of \cite{LiMa}). }
\caption{$\vartheta^2$ distribution of the combined 128.6\,h MAGIC observations between 2005 and 2007. $\vartheta^2$ is the squared angular distance between the source position and the reconstructed arrival direction of each event. The grey shaded area below the red crosses is the OFF data-sample, while the black crosses correspond to the ON data. The excess is point-like compared to the PSF with a significance of 7 standard deviations (calculated with formula 17 of Li \& Ma\ 1983). }
\label{fig:theta2}
\end{figure}

The $\vartheta^2$ distribution of the entire observing campaign is shown in Fig. \ref{fig:theta2}. An excess of 862 events over 12454 background events is apparent, corresponding to a significance of 7.0 $\sigma$ (calculated with formula 17 of Li \& Ma 1983).

\begin{figure}
\includegraphics[width=0.49\textwidth]{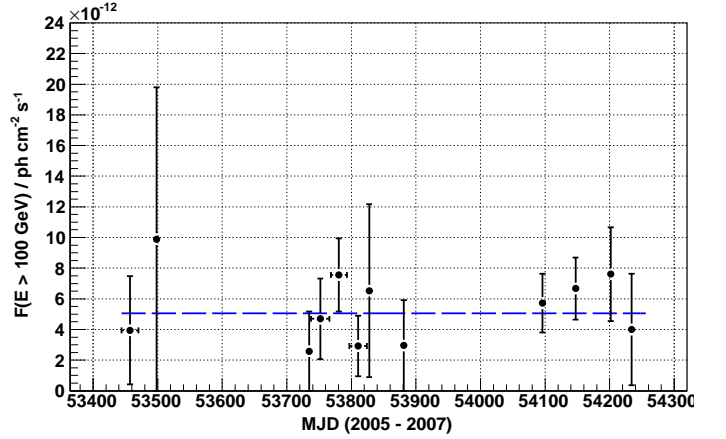}
\caption{Light curve of the integral $\gamma$-ray flux above 100\,GeV during the 2005--2007 MAGIC observing period. The dashed blue line corresponds to the fit result of a linear function to the data points, with a reduced $\chi^2/d.o.f. = 5.66/11$ and an average flux of $(5.06\pm0.77)\times 10^{-12}\,\mathrm{s}^{-1} \mathrm{cm}^{-2}$.}
\label{fig:LC}
\end{figure}

\begin{figure}
\includegraphics[width=0.49\textwidth]{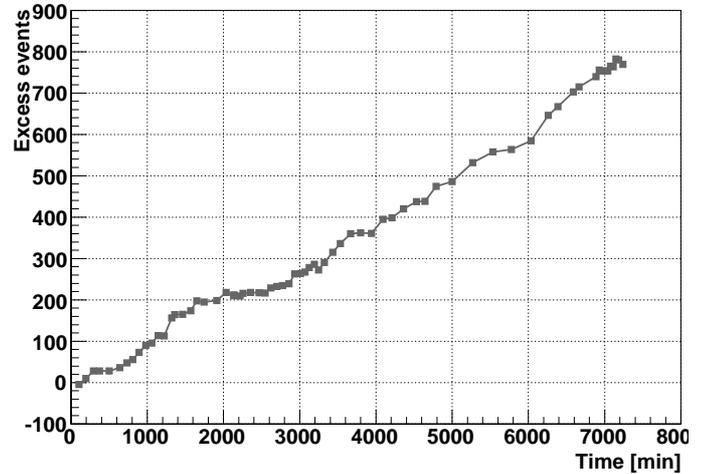}
\caption{Evolution of excess events from M87 over dead-time-corrected observing time. The linear increase is consistent with a constant flux of the source during the entire observing campaign. Only wobble data are shown in this figure in order to reduce systematic uncertainties in the ON-OFF subtraction.}
\label{fig:excess}
\end{figure}

We also investigated the stability of the signal over time. Since the observations are not evenly spaced (often interrupted by hardware changes of the telescope) and the length of each observing run differs greatly, the binning of the light curve is not uniform. Instead, we divided the data into calender months, starting from March 2005 until May 2007. Note that these bins have been chosen a posteriori and therefore no trial factors are involved. We assumed that each positive excess is related to M87 and derived the corresponding flux value based on the comparison with the Monte Carlo simulation of each period, taking into account the collection area, cut efficiency, trigger efficiency, and dead-time-corrected observing time.
As shown in the light curve of Fig. \ref{fig:LC}, the excess over the entire observing campaign is compatible with a fit to a constant flux of $(5.06\pm0.77)\times 10^{-12}\,\mathrm{s}^{-1} \mathrm{cm}^{-2}$ with a reduced $\chi^2$ of 0.51 (corresponding to a probability of 90\%). We also analysed each observing night independently, but none yielded a significant detection on its own. Fig. \ref{fig:excess} shows the cumulative excess distribution over time. The excess grows linearly, which is consistent with a constant emission, and shows that no significant flare occurred during our observing campaign. 

%Note that the data in 2005 are compatible within errors with the high flux state observed by H.E.S.S. \cite{HESSM87}. The reason is mainly a too short MAGIC observation time (and accordingly large error on the source flux) as well as the short time variability of the source (flares on a time scale of two days have been reported by the H.E.S.S. Collaboration).

\begin{figure}
\includegraphics[width=0.54\textwidth]{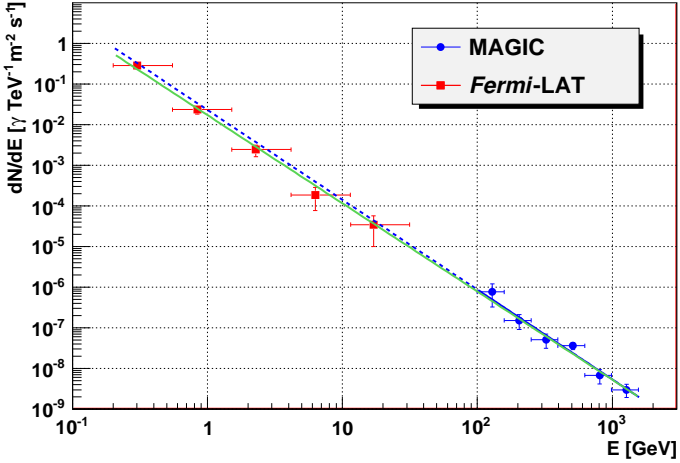}
\caption{Combined \textit{Fermi}--LAT and MAGIC differential energy spectrum over four orders of magnitude in energy starting from 100\,MeV until 2\,TeV. The \textit{Fermi}--LAT spectrum was obtained from \cite{Fermi}. The blue dashed line corresponds to the extrapolation of the fit to the MAGIC data points into the \textit{Fermi}--LAT energy range, while the green line represents the combined fit to the MAGIC and \textit{Fermi}--LAT data. No break or change in the spectral slope is apparent and both fits are statistically compatible. Note that the two observations are not contemporaneous, but \cite{Fermi} concluded that the source was most likely in a low-emission state during the observations.}
\label{fig:spec}
\end{figure}

A spectrum was derived using the same analysis technique as described for the extraction of the light curve. Additionally, a spill-over correction was applied to take into account the finite energy resolution (around 25\% for the data discussed here). Because M87 is a nearby source, the attenuation of VHE $\gamma$-rays by the extra-galactic background light is negligible in our energy range (Dom\'{\i}nguez et al.\ 2011).
The MAGIC spectrum is well described by a single power law: $dN/dE = f_0$ (E/300\,GeV)$^{\gamma}$ with a spectral index $\gamma=-2.21\pm0.21$ and a flux normalization  $f_0=(7.7\pm1.3) \times 10^{-8} \,\mathrm{TeV}^{-1} \mathrm{s}^{-1} \mathrm{m}^{-2}$. The MAGIC spectral points and fit are shown in blue in Fig. \ref{fig:spec}.

A detailed analysis of the VHE $\gamma$-ray variability of M87 is published elsewhere (Abramowski et al.\ 2012). The low state emission reported in our work is compatible to the flux observed by the H.E.S.S. Collaboration in 2004 and 2006, as well as the 2007 observations with the VERITAS array (see Fig. 1 in Abramowski et al.\ 2012).

The spectral index derived from our observations ($\gamma=-2.21\pm0.21$) is statistically compatible with previously reported results from H.E.S.S. ($\gamma$=2.62$\pm$0.35 for the 2004 data and $\gamma$=2.22$\pm$0.15 for 2005, Aharonian et al. 2006) as well as VERITAS ($\gamma$=2.31$\pm$0.17 in 2007, Acciari et al. 2008).

During the MAGIC observing periods no observations were performed in the GeV energy band. However, for comparison, we report (red squares, Fig. 4) the \textit{Fermi}--LAT spectrum obtained by integrating over the first ten months of all-sky survey data (Aug. 4, 2008 - May 31, 2009, Abdo et al.\ 2009). Based on contemporaneous X-ray and radio data, \cite{Fermi} argued that M87 was most likely in a low-emission state during the \textit{Fermi}--LAT observations. To investigate this hypothesis in more detail, we performed a combined fit to the MAGIC and \textit{Fermi}--LAT results (green line, Fig. 4). The combined fit yields a spectral index $\gamma=-2.17\pm0.03$ and a flux normalization at 300\,GeV $f_0=(7.1\pm1.0) \times 10^{-8} \,\mathrm{TeV}^{-1} \mathrm{s}^{-1} \mathrm{m}^{-2}$. The reduced $\chi^2$ changes from 1.14 to 0.86. The fit result is statistically compatible with the independent fit to the MAGIC data and confirms the low state during the \textit{Fermi}--LAT observations. There is no indication of a break or change in the spectral slope. We therefore included the \textit{Fermi}--LAT data in our modelling of the spectral energy distribution in the next section. It is worth noting that the peak of the spectral energy distribution appears to lie at particularly low energies, below 100\,MeV.

A detailed discussion of the potential systematics can be found in \cite{crab}. As a summary, the systematic energy scale error is about 16$\%$, the error of the flux normalization is estimated to be 11$\%$ and the systematic slope error is $\pm$0.2. Additionally, we have to consider how the combination of different observing modes and telescope configurations affects the systematic errors. We have combined Monte Carlo simulations that were tailored for each observing epoch (including the respective PSF) and tried to reconstruct the original Monte Carlo spectrum, which was fixed to a power law with a spectral index of -2.2 (identical to the detected spectral index of M87). From the maximum error in the reconstructed Monte Carlo spectrum we find that the error of the flux normalization should be $<$10$\%$ and the error of the slope should be $<$0.03 (note that this applies only to hard spectra with a spectral index $<$2.5, as is the case for the data discussed here). Finally, an independent second data analysis yielded compatible results.

\section{Modelling the SED}

\begin{figure}
\includegraphics[trim=14mm 15mm 0 0, clip, width=0.53\textwidth]{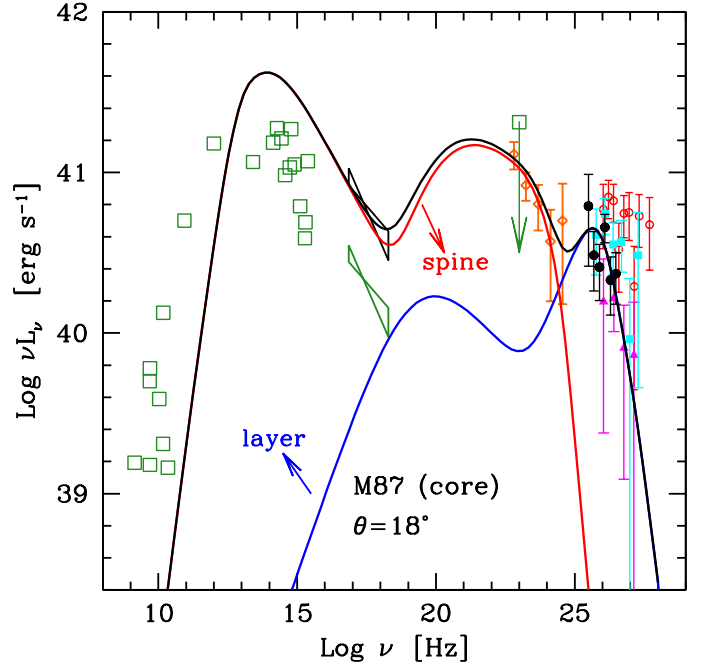}
\caption{SED of the core of M87 (green open squares) together with the MAGIC data points (black filled circles). For comparison we report the H.E.S.S. spectra taken in 2004 (magenta trangles) and 2005 (red open circles), from \cite{HESSM87}, and the VERITAS spectrum taken in 2007 (Acciari et al.\ 2008, cyan filled squares). The green bow-tie reports the X-ray spectrum as measured by Chandra in 2000 (from Balmaverde et al.\ 2006). As in \cite{tavecchio08} we reproduced the X-ray emission of the period 2005--2007 assuming the same slope and a higher normalization (black bow-tie). In the high-energy $\gamma$-ray range the green open square corresponds to the EGRET upper limit and the orange diamonds to the \textit{Fermi}--LAT energy spectrum from Abdo et al.\ 2009. The lines report the emission from the spine (red) and from the layer (blue), calculated with the parameters reported in Table 2, and their sum (black).}
\label{fig:mod}
\end{figure}

Fig. \ref{fig:mod} shows the SED of the core of M87, adapted from \cite{tavecchio08}. Green open symbols represent historical data, black filled circles show the MAGIC spectrum. For comparison we also show the H.E.S.S. spectra taken in 2004 and 2005 (magenta triangles and red open circles, respectively, from Aharonian et al.\ 2006) and the 2007 VERITAS spectrum (cyan squares, from Acciari et al.\ 2008). The 2005--2007 spectrum was very similar to that measured by H.E.S.S. in the ``low" 2004 state and by VERITAS in 2007. The green ``bow-tie" reports the nuclear X-ray spectrum measured by \textit{Chandra} in 2000 (Balmaverde et al.\ 2006). The black ``bow-tie" shows the average X-ray spectrum during the period covered by the MAGIC observations, obtained by scaling the spectrum from the year 2000 by a factor of four, as inferred from the light curve reported by \cite{Harris}. This procedure was necessary since no spectrum is available for the period of the MAGIC observations and both campaigns were not organised jointly, which results in very little overlapping observing time. The stability of the low-state spectrum supports this procedure of scaling. As explained in Section 3, we also included the non-contemporaneous \textit{Fermi}--LAT spectrum (orange open diamonds) obtained by integrating over the first ten months of all-sky survey data (Abdo et al.\ 2009). The SED shows two pronounced bumps, one peaking in the IR band, one extending from MeV to TeV energies.

As discussed in the introduction, models of the VHE emission from M87 include a class that considers the emission from ultra-relativistic electrons accelerated in the black hole magnetosphere in the vicinity of the horizon, and another group assuming emission from the jet. In the latter group some scenarios are based on variations of the general framework adopted for blazars, i.e. assuming that the emission comes from the innermost  regions of the jet, corresponding to the radio core (at 100--1000 Schwarzschild radii from the SMBH), while others propose that the radiation is produced at larger distances, possibly at a shock generated by a strong re-collimation of the flow. In the context of jet models, as discussed in \cite{tavecchio08}, one-zone models face a severe problem in describing the whole SED of M87, from radio up to TeV energies.
Indeed, the nuclear environment of M87 does not show evidence of significant sources of soft photons (e.g.\ from the torus or star formation sites) and therefore the inverse Compton emission is likely dominated by the SSC component. In that case, it is possible to derive the value of the required Doppler factor, $\delta$, which depends only on the synchrotron and SSC peak frequencies and luminosities and on the variability timescale. For an SED such as the one displayed by M87, whose synchrotron and SSC components peak in the IR and TeV band, respectively, one infers unacceptably high values of the Doppler factor, $\delta>100$.
The most direct solution to this problem is to assume that the jet emission region is structured, as in the ``decelerating jet" model of \cite{georg} or in the ``spine-layer" model of \cite{spinelayer}. Other possibilities include multiple regions moving into a wider jet, as expected from the inner regions before the initial collimation (Lenain et al.\ 2008), or emission by several randomly oriented active regions resulting from reconnection events in the jet (the so-called ``jets-in-the-jet" scenario of Giannios et al.\ 2009; Giannios et al.\ 2010). By construction, all these models predict that M87, if observed at smaller angles, would display an SED resembling those commonly observed for blazars.

We modelled the SED using the structured-jet model of \cite{spinelayer}, previously applied to M87 in \cite{tavecchio08} assuming that the low-state $\gamma$-ray emission is coming from the radio core, as is suspected for the high-state emission (Acciari et al.\ 2009). The model assumes that the jet has an inner fast core (spine) with a bulk Lorentz factor $\Gamma_{\rm S}$, surrounded by a slower layer, with a bulk Lorentz factor $\Gamma_{\rm L}$. In both regions, relativistic electrons emit through synchrotron and inverse Compton mechanisms. The velocity structure plays an important role in determining the emission properties of the jet. Indeed, the radiative interaction between the layer and the spine results in the amplification of the inverse Compton emission of both components. 
In the rest frame of each component the emission of the other is amplified because of the relative speed within the two regions: this "external" radiation contributes to the total energy density, enhancing the emitted inverse Compton radiation.  Depending on the parameters, this ``external Compton'' emission can dominate the internal synchrotron self--Compton (SSC) component that, especially in TeV blazars, is suppressed because scatterings mainly occur in the Klein--Nishina (KN) regime. We refer to \cite{spinelayer} and \cite{tavecchio08} for a full description of the model.

The model is completely specified by the following parameters: {\it i)} the spine is assumed to be a cylinder of radius $R$, height $H_{\rm S}$ (as measured in the spine frame) and in motion with bulk Lorentz factor $\Gamma_{\rm S}$; {\it ii)} the layer is modelled as a hollow cylinder with internal radius $R$, external radius $R_2$, height $H_{\rm L}$ (as measured in the frame of the layer), and bulk Lorentz factor $\Gamma_{\rm L}$. Each region contains a tangled magnetic field with intensity $B_{\rm S}$ and $B_{\rm L}$ respectively, and is filled by relativistic electrons assumed to follow a (purely phenomenological) smoothed broken power--law distribution extending from $\gamma_{\rm min}$ to $\gamma_{\rm max}$ with indices $n_1$, $n_2$ below and above the break at $\gamma_{\rm b}$.  The normalization of this distribution is calculated assuming that the system produces a (bolometric) synchrotron luminosity $L_{\rm syn}$ (as measured in the local frame), which is an input parameter of the model. We assume that $H_{\rm L} > H_{\rm S}$. As noted above, the seed photons for the IC scattering are not only those produced locally in the spine (layer), but we also consider the photons produced in the layer (spine).
The result of the modelling is reported in Fig. \ref{fig:mod}, where we show the SED of the emission produced by the spine (red) and the layer (blue) and their sum (black). The adopted parameters are reported in Table \ref{modelparam}.
The parameters we adopted are very similar to those derived in Tavecchio \& Ghisellini (2008) for the high 2005 state. In particular, we used the same values of the spine and layer dimensions and Doppler factors. Our model is therefore compatible with our initial assumption that the low- and high-state emissions are originating from the same emission region near the radio core. Since in the SED considered in Tavecchio \& Ghisellini (2008) there were no measures in the GeV band, the model of the flux in that band was unconstrained. The LAT data we used in the present model instead allow us to constrain the inverse Compton bump of the spine to a luminosity slightly lower than that assumed in Tavecchio \& Ghisellini (2008): this difference accounts for a stronger magnetic field derived here for the spine. Similarly to Tavecchio \& Ghisellini (2008), since we are forced to reproduce a high X-ray state, the model inevitably over-predicts the non-simultaneous flux in the IR-optical bands. However, as already noted in Tavecchio \& Ghisellini (2008), because the X-ray and optical fluxes are correlated (Perlman et al. 2003), we expect that high X-ray fluxes also corresponds to optical states higher than those presented in the SED.

\begin{table*}
\caption{Input parameters of the models for the layer and the spine shown in Fig. \ref{fig:mod}. All quantities (except the bulk Lorentz factors $\Gamma$ and the viewing angle $\theta$) are measured in the rest frame of the emitting plasma. The external radius of the layer is fixed to the value $R_2=1.2 \times R$. A detailed description of the model and its parameters can be found in the text.}
\label{modelparam}
\centering
\begin{tabular}{ c c c c c c c c c c c c c }
\hline \hline
& $R$    
& $H$  
& $L_{\rm syn}$  
& $B$  
& $\gamma_{\rm min} $  
& $\gamma_{\rm b} $ 
& $\gamma_{\rm max}$  
& $n_1$
& $n_2$
& $\Gamma$ 
& $\theta$ & \\
Emission Zone &cm  &cm &erg s$^{-1}$  &G & & & & & & &deg. & \\
\hline  
Spine  &7.5$\cdot10^{15}$ &3$\cdot10^{15}$ &4.7$\cdot10^{41}$ &2.1    &600 &2$\cdot10^{3}$ &1$\cdot10^{8}$ &2 &3.65 &12  &18 &\\ 
Layer  &7.5$\cdot10^{15}$ &6$\cdot10^{16}$ &1.6$\cdot10^{38}$ &0.35 &1   &2$\cdot10^{6}$ &1$\cdot10^{9}$ &2 &3.3 &4  &18  &\\
\hline
\end{tabular}
\end{table*}       

A feature of the ``spine-layer" model for M87 is that the IC emission from the layer, which accounts for the observed VHE emission, is partly absorbed through the interaction with the optical-IR photons produced in the spine and the subsequent production of electron positron pairs. This leads to a relatively soft spectrum, suitable for reproducing the low-level spectrum measured by MAGIC (which extends to $\sim$2\,TeV), but difficult to reconcile with the hard spectrum recorded during the high state by H.E.S.S. A possible solution of this problem is to enlarge the emission regions, since this would reduce the density of the target photons. However, the increase of the source size is limited by the observed short variability time-scales, $R< 5\times 10^{15} \delta$ cm. In summary, we are able to describe the VHE low-state emission with parameters similar to those adopted in \cite{tavecchio08} to reproduce the high state. We can therefore conclude that in the framework of the model adopted here, both high and quiescent states can originate from the same emitting region. The opacity of the emitting region to the $\gamma$-ray photons is a difficulty that also affects other models, especially those focusing on the acceleration in the BH magnetosphere, where the environment is expected to be extremely rich in IR photons that originate in the accretion flow (e.g.\ Neronov \& Aharonian 2007; Li et al.\ 2009). This problem, instead, could be relaxed in the ``mini-jets" model of \cite{giannios10}, due to the high Doppler factor $\delta\simeq 10$ assumed to characterize the emitting regions.
We also note that in the structured-jet scenario presented here, a strict correlation between these two spectral components is not required (even if it is possible), because the MeV--GeV emission originates in the spine while the VHE radiation is produced in the layer.

Another point concerns the jet power required to reproduce the SED. The inferred jet power (dominated by the Poynting flux associated to the magnetic field) is $P_{\rm jet}=1.5\times 10^{44}$ erg s$^{-1}$. The accretion luminosity in M87 is estimated to be about $L_{\rm accr}=10^{40}-10^{41}$ erg s$^{-1}$ (e.g.\ Perlman et al.\ 2001), which for a black hole mass of $M=6\times 10^9$ M$_{\odot}$ corresponds to Eddington ratios of $L_{\rm accr}/L_{\rm Edd}=10^{-8}-10^{-7}$. The luminosities as low as this, common among low-power radio galaxies are currently interpreted to be the result of a radiatively inefficient flow, with efficiencies of about $\eta\equiv L_{\rm accr}/{\dot M} c^2=10^{-4}-10^{-3}$ (e.g.\ Balmaverde et al.\ 2008). In that case the power request of the jet can be easily fulfilled by the accretion power, $\dot Mc^2 = L_{\rm accr}/\eta \sim 10^{45}$ erg s$^{-1}$. Finally, we also note that a magnetically dominated jet (Poynting to kinetic power ratio higher than $\sim$100) is postulated to power the ``mini-jets" in the \cite{giannios10} model.

A final comment concerns the angle within the plasma velocity and the line of sight, which in our model we assumed to be $\theta=18$ deg. This value is well within the range confidently derived through radio observations, $\theta=15-25$ deg (Acciari et al. 2009). Angles larger then $\sim 20$ deg would imply a de-boosting of the spine emission (see Fig. 2 of Tavecchio \& Ghisellini 2008), with the consequent increase in the {\it intrinsic} luminosity of the spine. The increased luminosity, in turn, would imply a higher power of the jet and, more importantly, would increase the optical depth to VHE photons emitted by the layer.

\section{Summary}

MAGIC has detected weak and steady VHE $\gamma$-ray emission from M87 between 2005 and 2007. The flux and spectral shape are consistent with a straight power law extrapolation of the published \textit{Fermi}--LAT spectrum at lower energies although both observations are not contemporaneous. Our measurements are also compatible with previously reported low states by VERITAS and H.E.S.S., which in turn suggests that the observed low-emission level and spectral characteristics are stable over a long time period.

We were able to describe this emission with a structured jet model, which separates the jet into a spine and an outer layer. The assumption that the low-state VHE $\gamma$-ray emission originates from the radio core of the jet like the high-state emission is consistent with our model. It should be noted, however, that the low-state emission alone cannot constrain the emission region due to the lack of variability of the measured signal and the softer VHE $\gamma$-ray spectrum. The parameters we derived from the model fit are consistent (even identical for the spine/layer dimensions and Doppler factors) with Tavecchio \& Ghisellini (2008) for the high 2005 state. Because the present model includes GeV data from the \textit{Fermi}--LAT, we were able to constrain the inverse Compton bump of the spine to a slightly lower luminosity compared to the assumptions in Tavecchio \& Ghisellini (2008), which results in a stronger magnetic field for the spine.

An interesting feature of the model is the transition region between the emission from the spine and that of the layer between 40 and 100\,GeV (the exact position depends on the chosen model parameters, see Fig. \ref{fig:mod} and Table 2). Currently, no measurements of M87 are available in this energy range and the non-contemporaneity of the multiwavelength data limits the accuracy of the spine-layer model. Since fall 2009, MAGIC consists of two 17\,m diameter telescopes, which observe in a stereoscopic mode. This upgrade has significantly improved the instrument's capabilities at energies below 100\,GeV as reported in \cite{Performance}. It is therefore possible that a future, deep observation of M87 with the stereoscopic MAGIC system and contemporaneous multiwavelength data will reveal a feature in the otherwise smooth power law spectrum of M87, confirming or disproving the validity of the structured jet model.

\begin{acknowledgements}
We would like to thank the Instituto de Astrof\'{\i}sica de
Canarias for the excellent working conditions at the
Observatorio del Roque de los Muchachos in La Palma.
The support of the German BMBF and MPG, the Italian INFN, 
the Swiss National Fund SNF, and the Spanish MICINN is 
gratefully acknowledged. This work was also supported by the CPAN CSD2007-00042 and MultiDark
CSD2009-00064 projects of the Spanish Consolider-Ingenio 2010
programme, by grant DO02-353 of the Bulgarian NSF, by grant 127740 of 
the Academy of Finland,
by the DFG Cluster of Excellence ``Origin and Structure of the 
Universe'', by the DFG Collaborative Research Centers SFB823/C4 and SFB876/C3,
and by the Polish MNiSzW grant 745/N-HESS-MAGIC/2010/0.
\end{acknowledgements}

\end{document}